\documentclass{article}
\usepackage{frascatiphys,here,graphicx,subfigure}

\newcommand{\be}{\begin{equation}}
\newcommand{\ee}{\end{equation}}
\newcommand{\bea}{\begin{eqnarray}}
\newcommand{\eea}{\end{eqnarray}}

\renewcommand{\sc}{\slashchar}

\def\slashchar#1{\setbox0=\hbox{$#1$}           
   \dimen0=\wd0                                 
   \setbox1=\hbox{/} \dimen1=\wd1               
   \ifdim\dimen0>\dimen1                        
      \rlap{\hbox to \dimen0{\hfil/\hfil}}      
      #1                                        
   \else                                        
      \rlap{\hbox to \dimen1{\hfil$#1$\hfil}}   
      /                                         
         \fi}                                         %

\begin{document}

\title{LIGHT HADRON SPECTRUM IN THE INSTANTON LIQUID MODEL}
\author{
P. Faccioli \\
{\em Dipartimento di Fisica, Universit\`a degli
Studi di Trento and I.N.F.N.}\\
M. Cristoforetti,\\
{\em Physik Department, Technische Universit\"at M\"unchen}\\
M.C. Tichy, \\ 
{\em Insitut f\"ur Theoretische Physik, Universit\"at T\"ubingen},\\
{\em and Dipartimento di Fisica, Universit\`a degli Studi di Trento}\\
M.Traini, \\
{\em and Dipartimento di Fisica, Universit\`a degli Studi di Trento, I.N.F.N. 
and E.C.T.*}
}

\maketitle
\baselineskip=11.6pt

\begin{abstract}
We  review our recent study of the role played by the chiral interactions induced by instantons, in  the
lowest-lying sector of the light hadron spectrum. 
We discuss how the ordering of the lowest meson
and baryon excitations is explained by the structure of the instanton-induced quark-quark
and gluon-gluon interaction. We focus on  the pion, nucleon, 
vector- and axial-vector mesons, and on the scalar glueball. 
We find that all these hadrons are bound in this model and have realistic masses.
\end{abstract}

\baselineskip=14pt

\section{Introduction}

The  spectrum of the lightest hadrons encodes the  information about the way the $u-$ and $d-$ quarks interact with gluons, 
at different distance scales. The $\sim 400$~MeV splitting between parity partners, e.g. vector- and axial-vector- 
mesons, suggests that the interactions associated with the spontaneous breaking of chiral symmetry are very important in the 
low-lying sector of the  spectrum.
Similarly, the large splitting between the pion and the $\eta'$ implies that topological interactions related to the axial anomaly are giving significant contribution. On the other hand, splitting between parity-partners is much reduced for higher resonances, 
and there have been claims that chiral symmetry may even be restored, up in the spectrum~\cite{glozman}.

At a qualitative level, the large contribution of the chiral forces to the internal dynamics of the lowest-lying hadrons can be explained as follows. The wave-function of the ground-state hadrons and of the lowest resonances is narrower than that of the higher excitations. 
Consequently, quarks in low-lying states and low-lying resonances are on average closer to each other and therefore 
relatively less sensitive  to the very long-distance, confining part of the quark-quark interaction. 
On the other hand, they are very sensitive to the non-perturbative correlations which take place at the intermediate distance 
scales, $\sim 1/\Lambda_\chi\sim 0.2-0.3$~fm, where $\Lambda_\chi$ is the scale associated to chiral symmetry breaking.     
Conversely, up in the spectrum, the hadron wave-function extends for larger distances and quarks begin to experience the effect of 
the confining forces, which take place at the QCD scale $1/\Lambda_{QCD}\sim$~1~fm.

Within such a scenario, two questions emerge naturally: (i) what is the microscopic origin of the interactions associated with the spontaneous breaking of chiral symmetry?  (ii) are any of the light hadrons completely dominated by such chiral forces, 
to a point that they would exist even if confining correlations were completely removed?
In this talk, I will review our recent attempts to address these questions in the context of the 
Interacting Instanton Liquid Model (IILM).

It has long been argued that instantons of size $\sim~0.3$~fm represent the main vacuum gauge field configurations responsible 
for the  non-perturbative dynamics at the chiral scale~\cite{shuryak82}. Recent lattice  studies~\cite{gattringer} have provided strong evidence in support of such an hypothesis. As a consequence of the spontaneous breaking of chiral symmetry, 
quark propagating in the instanton vacuum develop an effective mass of $\sim 350-400$~MeV, 
hence this model provides a connection between current and constituent quarks.

The main drawback of the instanton models is that they do not provide confinement. On the one hand, 
this is a serious problem. It implies that instantons cannot be the only important non-perturbative gauge field fluctuations in the QCD vacuum. On the other hand, just because of the lack of confinement, the instanton models represent a convenient framework in which the effect of  chiral symmetry breaking can be isolated and the questions listed above can be addressed. 

\section{The instanton-induced interaction in the different hadrons}

The  interaction generated by instantons is not equally important in all hadrons. 
Due to the specific quantum-number structure of the 't~Hooft vertex,  and of the instanton gauge field 
there are channels in which the instanton effect come at leading order in the instanton vacuum diluteness, $\kappa\sim0.1$ 
and channels in which they come at sub-leading orders. This feature of the model is very important as it provides a natural explanation to  a number of observed  phenomena~\cite{hadronsILM}: for example, it explains the well-known $\Delta~I=1/2$ rule for non-leptonic hyperon and kaon decays, or the very early on-set of the perturbative regime in the $\gamma \gamma^*\to \pi^0$ transition form factor, relative to the strongly non-pertrubative behavior of the space-like pion form factor~\cite{pionFF}.   

As far as the hadron spectrum is concerned, the structure of the instanton-induced interaction correctly accounts for the ordering of the  lowest-lying states, i.e. where we expect chiral forces to be important.  In fact, leading instanton forces are most attractive in the pion, but are suppressed in the $\rho$-meson and $A_1$-meson and are even repulsive in the $\eta'$-meson.
Similarly, they are very strong and attractive in the nucleon and are suppressed in the $\Delta$-isobar.
Remarkably, the same dynamical mechanism can explain also the ordering of the lightest glueball excitations observed in lattice QCD simulations, with strong attraction in the scalar channel, suppression in the tensor channel and repulsion in the speudo-scalar channel.

\begin{figure}[t!]
        {\includegraphics[scale=0.34]{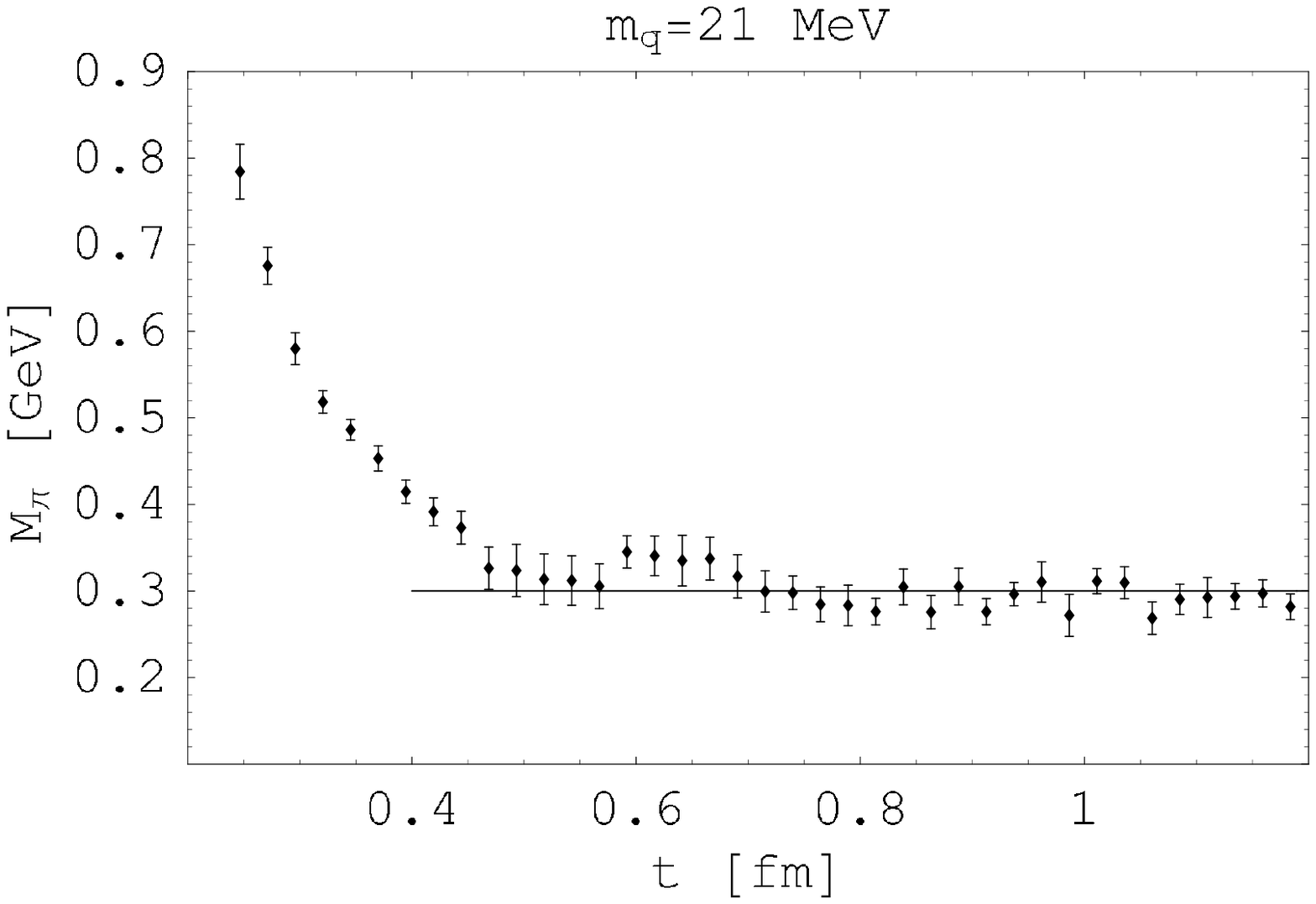}}
        {\includegraphics[scale=0.34]{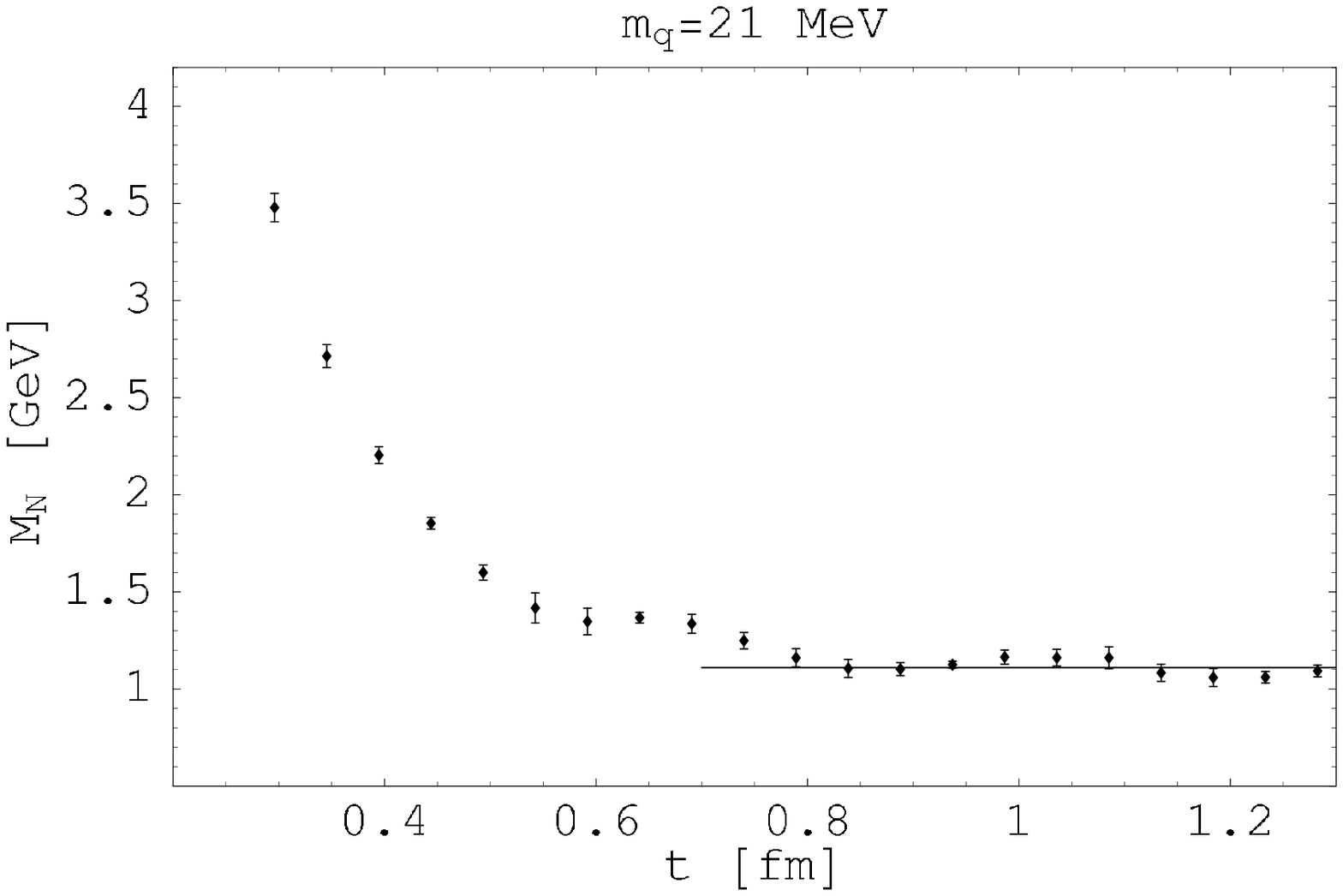}}
\begin{center}
  \caption{Typical effective mass plots obtained in the IILM and used to extract the pion (left panel) and nucleon (right panel) masses.}
\end{center}
\label{pionnucleon}
\end{figure}

\section{Hadron Mass Calculation in the IILM}

In the IILM, the QCD path integral over all possible gluon field configurations is replaced by 
\begin{equation}
\mathcal{Z}_{QCD}\simeq\mathcal{Z}_{ILM} =\sum_{N_+, N_-} \frac{1}{N_+!N_-!}\int\prod_i^{N_++N_-}\textrm{d}\Omega_id(\rho_i)e^{-S_{int}}\prod_i^{N_f}\textrm{det}(i\sc{D}+im_f)\nonumber
\end{equation}
Here, $\textrm{d}\Omega_i=\textrm{d}U_i\textrm{d}^4z_i\textrm{d}\rho_i$ is the measure in the space of collective coordinates, color orientation, position and size, associated with the single instantons. Quantum fluctuations are included in Gaussian approximation, through the semi-classical instanton amplitude $d(\rho_i)$. $S_{int}$ is a bosonic  interaction between pseudo-particles which includes a  long-distance attractive interaction derived semi-classically and a  short-range repulsive core, introduced phenomenologically in order to remove large-sized instantons from the vacuum. In the formulation of the model we have considered~\cite{chiralILM}, the strength of such a repulsion is the only phenomenological parameter, which has to be tuned to reproduce observations.

In a field-theoretic framework, the information about the hadron spectrum is encoded in the two-point Euclidean correlation functions,
\be
G_{H}(\tau)=\int\textrm{d}^3\underline{x}\langle0|T[j_H(\underline{x},\tau)\overline{j}_H(\underline{0},0)|0\rangle,
\ee
where $J_H$ is an operator which excites states of hadrons with the quantum numbers of the hadron $H$. Once such a correlation function has been evaluated, the mass of lowest-lying stable hadron can then be extracted from the plateau in the large Euclidean time limit of the effective mass, i.e. using 
\be
M_{H}=\lim_{\tau\rightarrow\infty}M_{H}^{eff}(\tau)\qquad M_{H}^{eff}(\tau)=\frac{1}{\Delta \tau}\ln\frac{G_{H}(\tau)}{G_{H}(\tau+\Delta\tau)}.\nonumber
\ee
\begin{figure}[t!]
        {\includegraphics[scale=0.36]{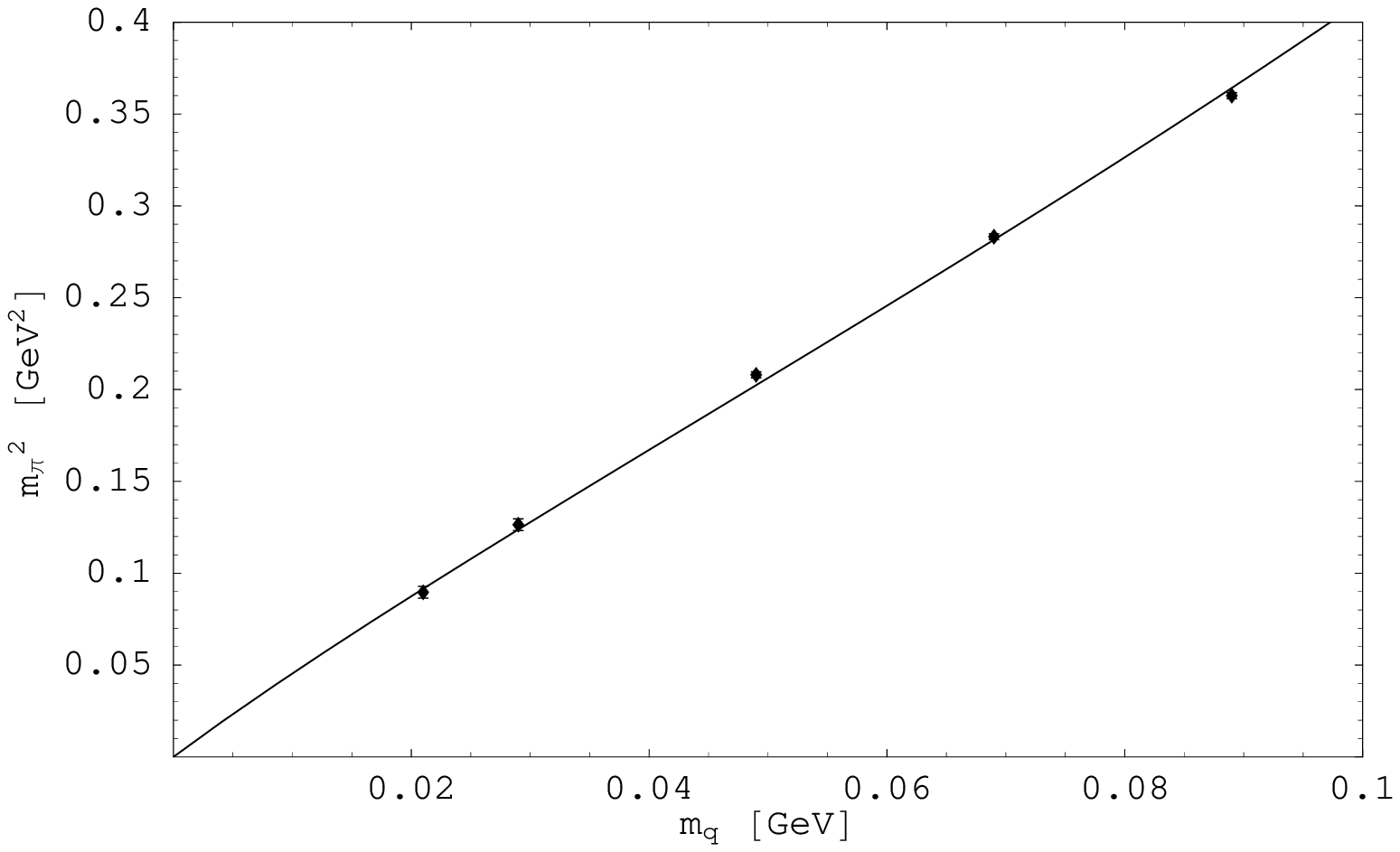}}
        {\includegraphics[scale=0.2]{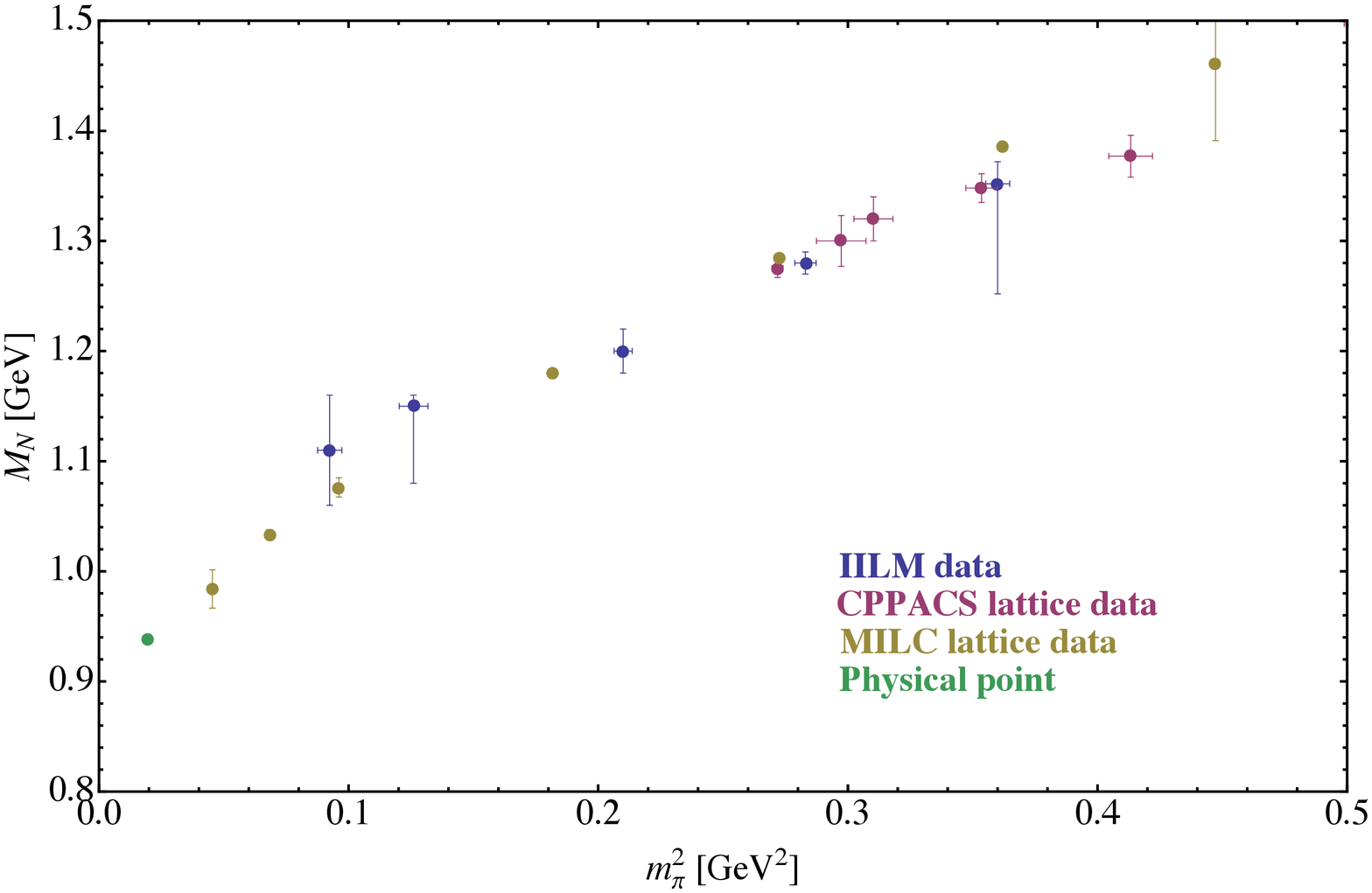}}
\begin{center}
  \caption{Left Panel: The pion mass as a function of the quark mass obtained in the IILM and compared with the extrapolation formula derived from chiral perturbation theory. Right Panel: The nucleon mass as a function of the pion mass squared obtained in IILM and in lattice calculations.}
\end{center}
\label{exfig}
\end{figure}
The pion and the nucleon are the two hadrons in which the instanton-induced interaction is most intense. Typical  effective mass plots obtained from IILM calculations~\cite{chiralILM} are shown in Fig.1 and are qualitatively indistinguishable from those obtained in lattice QCD simulations. They clearly display a plateau, which is the signature for the existence of a bound-state. 
We have extracted the hadron masses corresponding to  five different values of the quark mass in the range $20<m_q<90$ MeV. 
The behavior of the pion (nucleon) mass with $m_q$ ($m_\pi^2$) is presented in Fig.~2. We have fitted the chiral behavior of the pion mass on the quark mass using the extrapolation formula obtained to $\mathcal{O}(p^2)$ from chiral perturbation theory. This leads to low-energy effective coefficients $f_0=85~$MeV and $\langle \bar q q\rangle = (-259 \textrm{MeV})^3$, in excellent agreement with 
phenomenology.   On the other hand, the calculated nucleon masses at different values of the pion mass agree 
very well with the available results of lattice QCD simulations.

\begin{figure}[t!]
        {\includegraphics[scale=0.35]{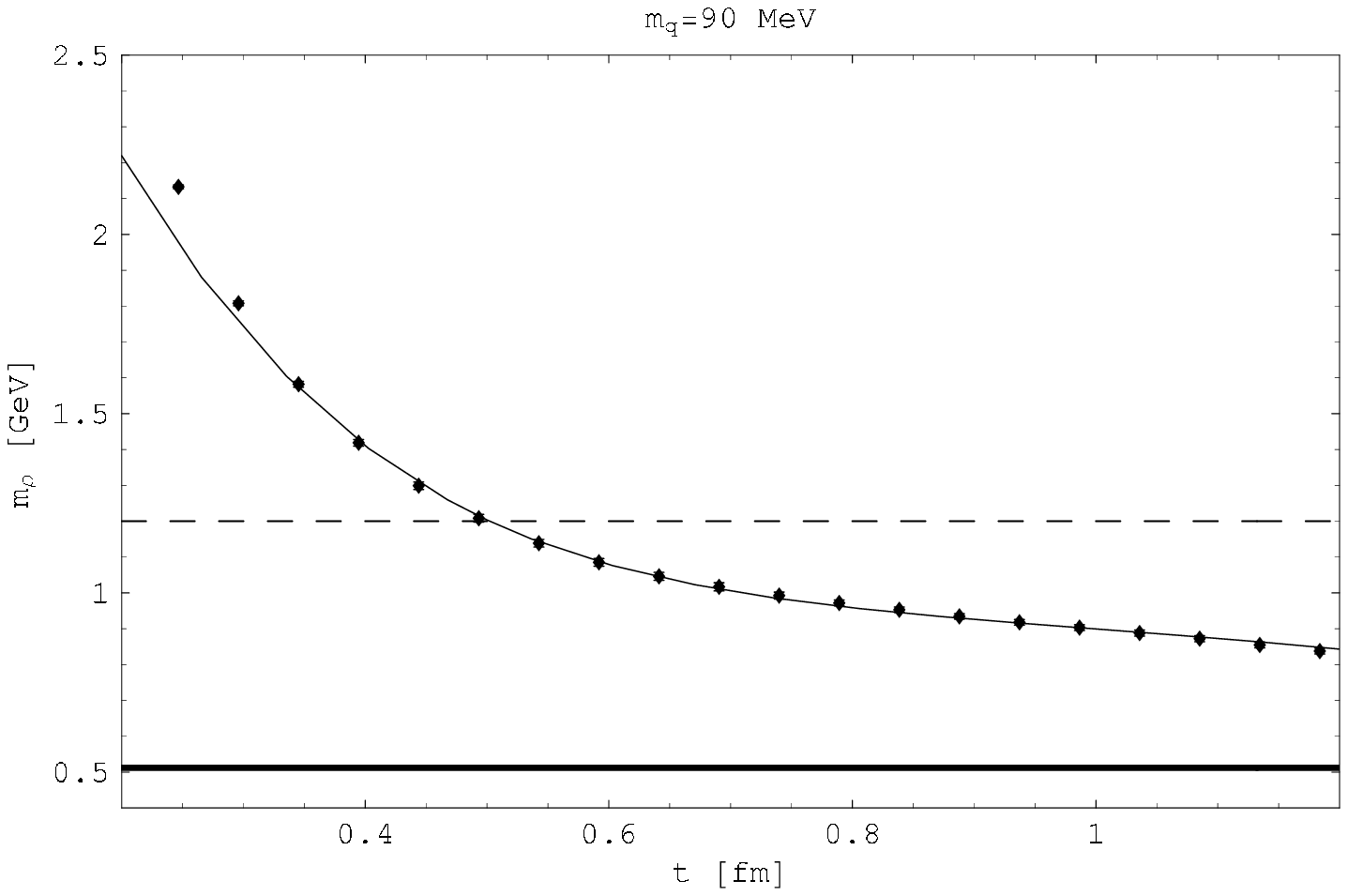}}\hspace{0.5cm}
        {\includegraphics[scale=0.39]{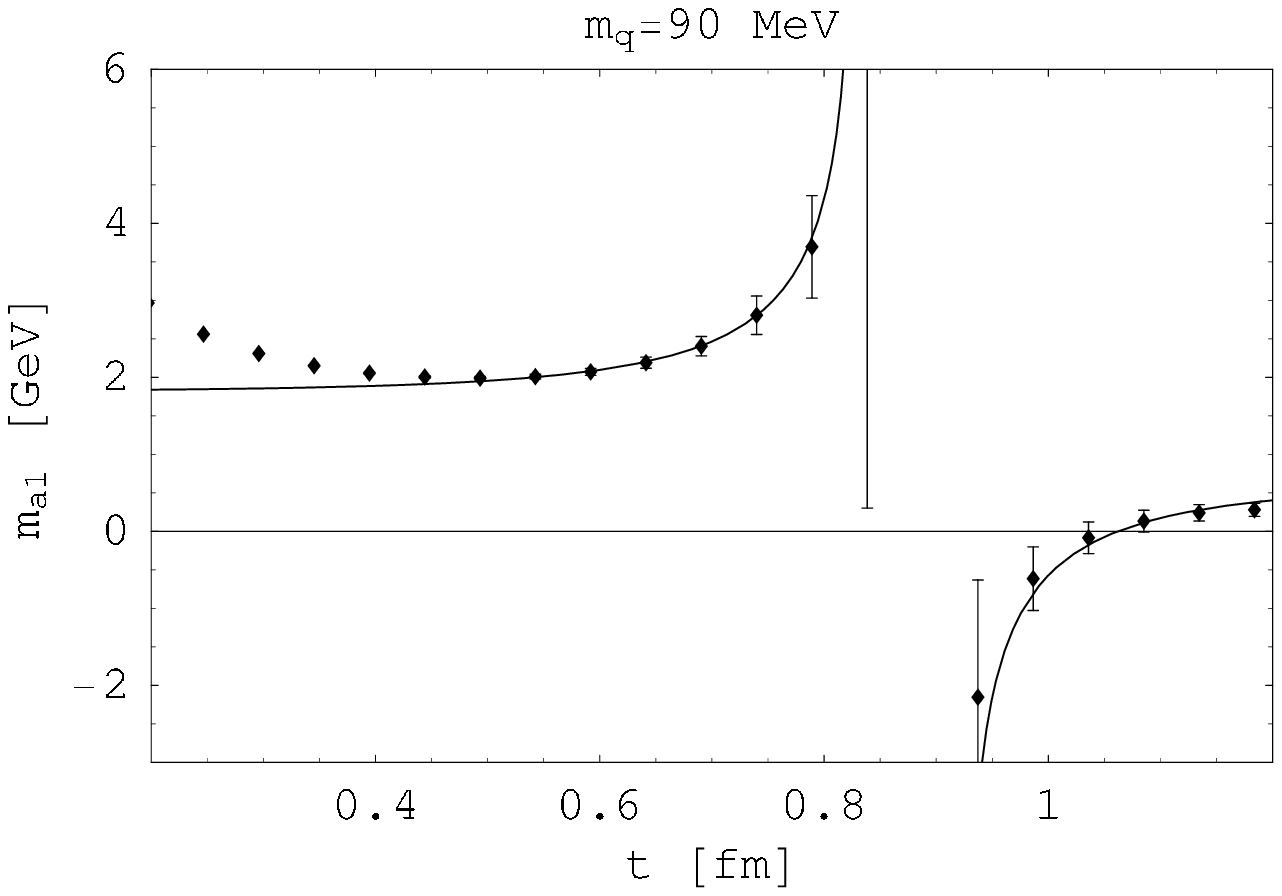}}
\begin{center}
  \caption{The effective mass plot for the $\rho$-meson (left panel) and $A_1$ meson (right panel) evaluated in the IILM (points) and compared with the behavior expected from the structure of the spectral function (line). The significance of the thick line and 
  of the dashed line is explained in the original paper~\cite{vectorILM}}
\end{center}
\label{resonances}
\end{figure}

Extracting information about the mass of the unstable vector and axial-vector meson resonances from the effective mass plot analysis is much harder than in the case of  ground-state hadrons. In fact, if the quark mass is sufficiently small,  the effective mass does not converge to the mass of the lowest resonance, but to the invariant mass of the decay products, at threshold. In order to be able to extract the masses of $\rho$ and $A_1$ mesons from IILM correlations functions, the expected specific functional form of their effective mass  was investigated in detail in \cite{vectorILM}, by using the experimental information about their spectral function, available from ALEPH. In Fig.~\ref{resonances} we compare the expected shape of the effective mass (line) with the points obtained in the IILM and find  agreement. Note that the singularity in the axial-vector channel arises from the interference of the pion and axial-vector contributions and  therefore represent a clean evidence that both such states exist in the instanton vacuum.

On the other hand, we have found that the calculated $\rho$- and $A_1$- meson mass are almost $30\%$ larger than
the experimental value. This fact can be interpreted as a signature that, in such hadrons, confinement begins to play an significant role.

We conclude this section by mentioning our recent calculation of the mass of the scalar glueball, in the instanon vacuum\cite{glueballILM}. 
While numerical calculations completely analog to the ones performed for the nucleon and pion are in presently progress, here we discuss our recent results based on the Single Instanton Approximation (SIA)~\cite{sia}.

The SIA  follows from the observation that if the instanton liquid is  sufficiently dilute, short-sized correlation functions are saturated by the contribution of a {\it single} pseudo-particle in the ensemble, the one closest to the endpoints  of the correlator.

The main advantage of the SIA is that it allows to obtain predictions from analytic calculations, rather than from Monte Carlo numerical simulations. The prize to pay is that the SIA can be used to compute  correlation functions with external momenta of the order of several GeV 
and for Euclidean times smaller than $\sim 1$~fm.

In order to reliably use the SIA, it is convenient to introduce a {\it momentum-dependent effective mass}, 
\be
M_{eff}(\tau,{\bf p}) = \sqrt{E^2_{eff}(\tau,{\bf p}) - {\bf p}^2}, \qquad E_{eff}(\tau,{\bf p})= -\frac{d}{d\tau} \log G_S(\tau, {\bf p}). 
\label{pMeff}
\ee
It is straightforward to show that, if the lowest scalar glueball excitation in the spectrum is a single-particle bound-state, then in the large Euclidean time limit $M_{eff}(\tau,{\bf p})$ must stop depending on $\tau$ and on ${\bf p}$ and converge to the glueball's mass:
$ \lim_{\tau\to\infty} M_{eff}({\tau,\bf p}) =  M_{0^{++}}.$

\begin{figure}[t!]
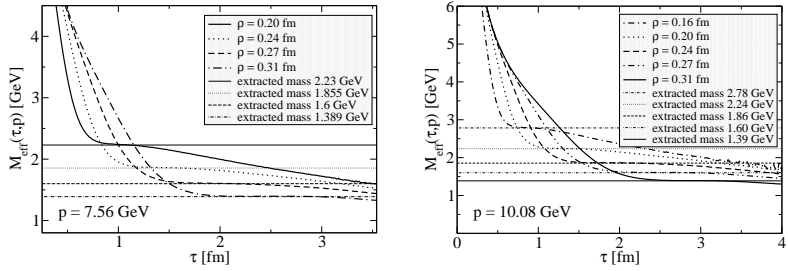

        {\includegraphics[scale=0.2]{plot-pp30.eps}}\hspace{0.5cm}
        {\includegraphics[scale=0.2]{plot-pp40.eps}}
\begin{center}
  \caption{The momentum-dependent effective mass plot for the scalar glueball mass evaluated at for several
average instanton sizes and two different momenta.}
\end{center}
\label{resultsSIA}
\end{figure}

Results for the SIA momentum-dependent effective mass $M_{eff}(\tau,{\bf p})$ are reported in 
Fig.~\ref{resultsSIA}, which  shows how the effective mass plot calculated at two different momenta, in a range of different average instanton sizes.
These plots clearly show that there exists a range of Euclidean times for which the momentum-dependent effective mass becomes independent on both  momentum and Euclidean time. The scalar glueball mass predicted by the model is $M_{O^{++}}=1.290-1.420$~GeV, in good agreement with the recent results of lattice calculations of Meyer and Teper~\cite{mayer}~$M^{latt.}_{0^{++}}=1475(30)_{stat.}(65)_{sys.}$~MeV. At very large times, the SIA breaks down and the correlators start depending on $\tau$ again.

\section{Conclusions}
In this talk, we have reviewed our recent attempts  to use the IILM to investigate the structure of the lowest-lying part of the hadron spectrum. We have found that nucleon, pion, vector- and axial- vector mesons as well as the lightest scalar glueball can be bound and have realistic masses, even in the absence of confinement. These results complement previous studies, in which it was shown that also the electro-weak structure of light hadrons can be well understood in this model\cite{pionFF,nucleonILM}.

\end{document}